\documentclass[conference]{IEEEtran}
\IEEEoverridecommandlockouts
\usepackage{cite}
\usepackage{amsmath,amssymb,amsfonts}
\usepackage{algorithmic}
\usepackage{graphicx}
\usepackage{textcomp}
\usepackage{xcolor}
\usepackage{hyperref}
\def\BibTeX{{\rm B\kern-.05em{\sc i\kern-.025em b}\kern-.08em
    T\kern-.1667em\lower.7ex\hbox{E}\kern-.125emX}}
\begin{document}
\newcommand{\Ted}[1]{{\it\small\textcolor{blue}{[ {#1}\ --Ted ]}}}
\title{GAEI-UNet: Global Attention and Elastic Interaction U-Net for Vessel Image Segmentation\\
}

\author{\IEEEauthorblockN{1\textsuperscript{st} Ruiqiang Xiao}
\IEEEauthorblockA{\textit{School of Science} \\
\textit{Hong Kong University of Science and Technology}\\
Hong Kong, Hong Kong SAR \\
rxiaoad@connect.ust.hk}
\and
\IEEEauthorblockN{2\textsuperscript{nd} Zhuoyue Wan}
\IEEEauthorblockA{\textit{School of Science} \\
\textit{Hong Kong University of Science and Technology}\\
Hong Kong, Hong Kong SAR \\
zwanah@connect.ust.hk}
\and
\IEEEauthorblockN{3\textsuperscript{rd} Yang Xiang}
\IEEEauthorblockA{\textit{Department of Mathematics} \\
\textit{Hong Kong University of Science and Technology}\\
Hong Kong, Hong Kong SAR \\
maxiang@ust.hk}
}
\author{
	\IEEEauthorblockN{
		Ruiqiang Xiao\IEEEauthorrefmark{1}, 
		Zhuoyue Wan\IEEEauthorrefmark{2}
	\IEEEauthorblockA{\IEEEauthorrefmark{1}\textit{School of Science}\\ \textit{Hong Kong University of Science and Technology}\\ Email: rxiaoad@connect.ust.hk}
	\IEEEauthorblockA{\IEEEauthorrefmark{2}\textit{School of Science}\\ \textit{Hong Kong University of Science and Technology}\\ Email: zwanah@connect.ust.hk}
 }
} 

\maketitle

\begin{abstract}
Vessel image segmentation plays a pivotal role in medical diagnostics, aiding in the early detection and treatment of vascular diseases. While segmentation based on deep learning has shown promising results,  effectively segmenting small structures and maintaining connectivity between them remains challenging. To address these limitations, we propose GAEI-UNet, a novel model that combines global attention and elastic interaction-based techniques. GAEI-UNet leverages global spatial and channel context information to enhance high-level semantic understanding within the U-Net architecture, enabling precise segmentation of small vessels. Additionally, we adopt an elastic interaction-based loss function to improve connectivity among these fine structures. By capturing the forces generated by misalignment between target and predicted shapes, our model effectively learns to preserve the correct topology of vessel networks. Evaluation on retinal vessel dataset -- DRIVE demonstrates the superior performance of GAEI-UNet in terms of SE and connectivity of small structures, without significantly increasing computational complexity. This research aims to advance the field of vessel image segmentation, providing more accurate and reliable diagnostic tools for the medical community. The implementation code is available on \href{https://github.com/keeplearning-again/GAEI-UNet}{\textit{Code}}.
\end{abstract}


\section{Introduction}
Medical image analysis plays a crucial role in medicine by providing valuable information for efficient diagnostic and treatment processes for radiologists and clinicians \cite{guo2014microfluidic}. One important area of medical image analysis is the segmentation of retinal vessels, which is useful for diagnosing diseases such as hypertensive retinopathy. Medical imaging devices such as X-ray, CT, and MRI can provide detailed anatomic and functional information about diseases and abnormalities inside the body \cite{shen2016learning}\cite{song2015lung}\cite{lee2001automated}\cite{heimann2007statistical}

\begin{figure}[htbp]
    \centering
    \includegraphics[width=0.3 \textwidth]{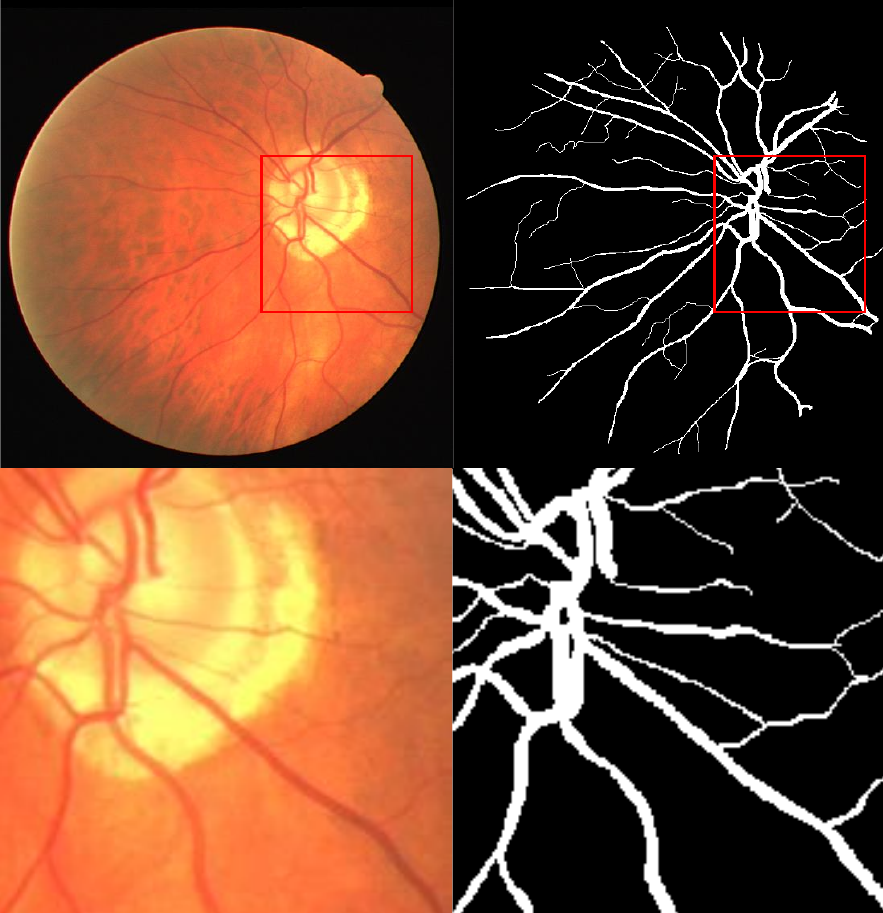}
    \caption{Vessel images and the corresponding ground truths in the
public dataset. The first and second columns are retinal vessel images
from the DRIVE dataset. The small vessel structure takes up a large proportion of the segmentation task.} 
    \label{fig0}
\end{figure}

The objective of image segmentation is to divide an image into distinct and non-overlapping regions based on either intensity or textural information \cite{acelastic}.
In medical imaging, the complexity of the background and foreground relationships, occlusions, lighting conditions, and small structures can all have an impact on the final diagnostic results, particularly in the context of medical image segmentation \cite{liu2021review}. For instance, in the case of Fig. \ref{fig0}, the complex tree-like structure of the retinal blood vessels, which consist of numerous small and fragile vessels that are closely connected, presents a significant challenge \cite{ricci2007retinal}\cite{moccia2018blood}. Additionally, the lack of clear contrast between the blood vessel area and the background, as well as the susceptibility of fundus images to uneven lighting and noise, further complicates the task of retinal blood vessel segmentation \cite{liu2021review}. Therefore retinal blood vessel segmentation remains a challenging task.

\begin{figure*}[htbp]
    \centering
    \includegraphics[width=0.85 \textwidth]{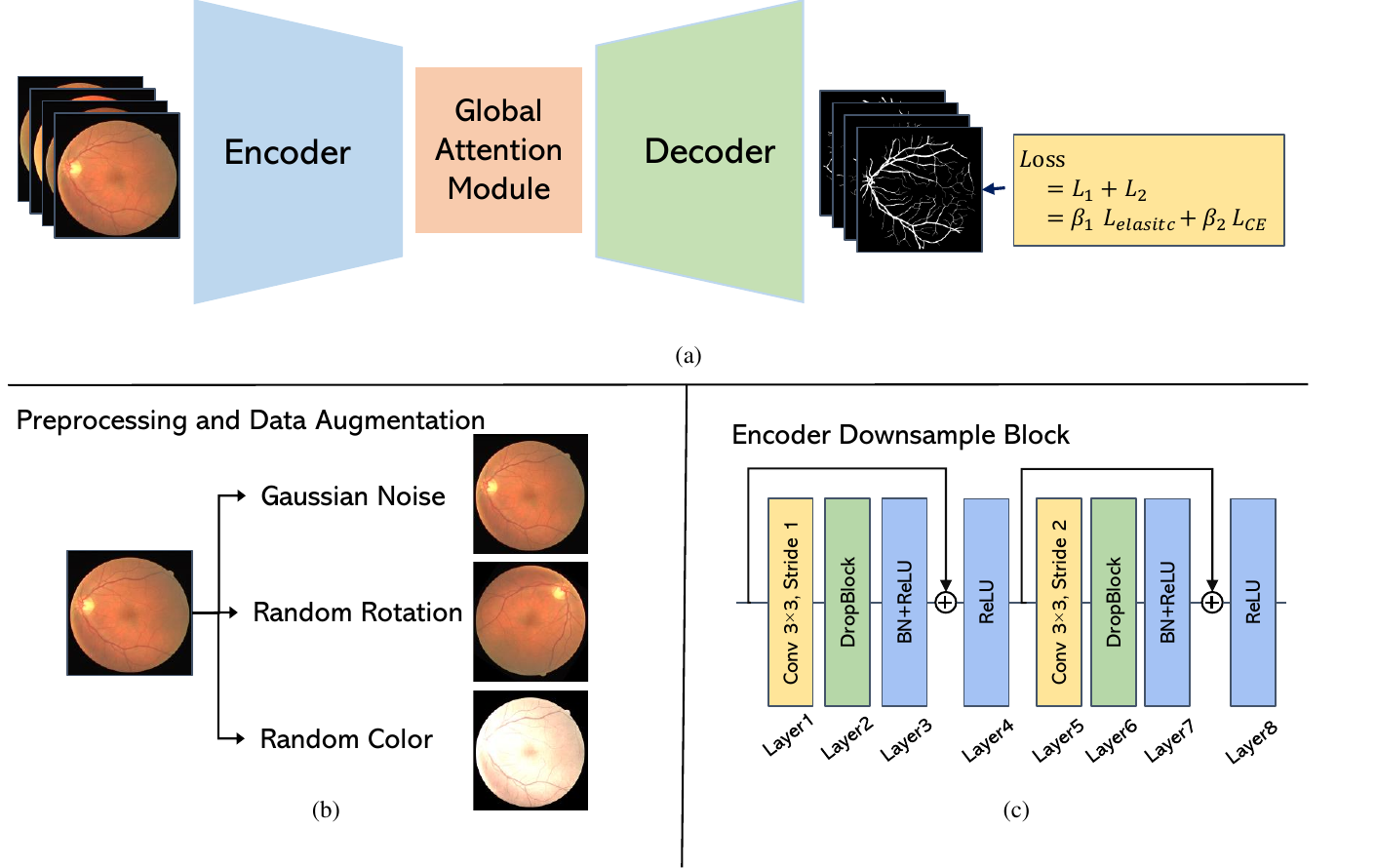}
    \caption{(a)Diagram of the proposed GAEI-UNet. The whole structure contains U-shape encoder-decoder architecture. The global attention module is added as the enhancement of the feature formulated after the encoder part. (b)The data augmentation combines Gaussian noisy process, random rotation process, and random color process. (c)The modified residual block, as well as the encoder downsample block, adds a dropout layer before each BatchNorm layer to reduce overfitting and adds a skip connection to acquire non-linear relations.} 
    \label{fig1}
\end{figure*}

Historically, medical image segmentation has relied on edge detection and template matching techniques \cite{lee2001automated}. For instance, circular or elliptical Hough transforms have been employed in optic disc segmentation \cite{zhu2008detection}\cite{aquino2010detecting}, while template matching has been utilized for spleen segmentation in MRI sequence images \cite{mihaylova2018spleen} and ventricular segmentation in brain CT images \cite{chen2009automated}.
Deformable models have also been proposed for medical image segmentation. For instance, a shape-based method utilizing level sets has been proposed for both two-dimensional segmentation of cardiac MRI images and three-dimensional segmentation of prostate MRI images \cite{tsai2003shape}. 
One of the main drawbacks of these methods is their reliance on hand-crafted features to achieve segmentation results. Firstly, it can be challenging to design representative features suitable for different applications. Secondly, the features that work well for one type of image may not be effective for another type. As a result, there is a lack of a general approach to feature extraction.

With the rapid development of convolutional neural networks in image processing, deep learning has been used for medical image segmentation based on that architecture\cite{ATLI2021271}, such as optic disc segmentation, blood vessel detection, lung segmentation, cell segmentation, etc. U-Net \cite{unet} is a common and well-known backbone network. U-Net consists of a typical downsampling encoder and upsampling decoder structure, with a "skip connection" between them. It combines local and global context information through the encoding and decoding process. Due to the excellent performance of U-Net, many recent methods for retinal blood vessel segmentation are based on U-Net. For instance, Wang et al. \cite{wang2019dual} reported the Dual Encoding U-Net (DEU-Net), which remarkably enhances the network's capability of segmenting retinal vessels in an end-to-end and pixel-to-pixel way. Wu et al. \cite{wu2019vessel} proposed Vessel-Net, which uses a strategy that combines the advantages of the initial method and the residual method to perform retinal vessel segmentation. Zhang et al. \cite{zhang2019attention} proposed AG-Net, which designed an attention mechanism called "Attention Guide Filter" to better retain structural information. Although these U-Net variants perform well, they may still struggle to achieve good segmentation results in complex scenes or when fine-grained segmentation is required. This is often attributed to the network's inability to extract sufficient and effective features\cite{su2022yolo}, such as non-local features at different scales and distances.

Moreover, the commonly used loss functions in the deep segmentation task are pixel-wise loss functions. For instance, loss functions such as cross-entropy (CE) \cite{zhou2017focal} and dice coefficient (DC) \cite{unet} can guide deep neural networks to focus on extracted features from specific regions \cite{tran2016fully}. While this can result in good classification and segmentation performance, low resultant loss function values may not necessarily correspond to a meaningful segmentation. For example, a noisy result can add many contours in the background, representing incorrect segmentation, and object boundaries can be fuzzy due to the difficulty of classifying pixels near the boundary. So there should be a more precise loss evaluation method to consider the topological or physical properties of the target objects rather than learning each pixel individually.

In this study, in parallel with improvements in the capabilities of those U-net-based models, we have discovered some techniques to improve segmentation performance. we propose a novel model called global attention and elastic interaction U-net (GAEI-UNet). We tested the performance of our method on the DRIVE dataset and compared the accuracies with results from other networks. Experiment results show that GAEI-UNet has leading accuracy.

The contributions of this paper can be summarized as follows:
\begin{itemize}
    \item We apply U-Net as the encoder with the modified residual dropout block for blood vessel segmentation to promote the capability to avoid gradient vanishing and overfitting.
    \item By leveraging an improved attention mechanism, an Attention-based Global context aggregator (AGCA) is proposed to enhance the extraction of high-level semantic features.
    \item We combine long-range elastic interaction-based loss function (EIL) with Cross-entropy Loss to capture global, long-range information for the long, thin structures in medical images. 
\end{itemize}

\section{Method}

\subsection{Network Architecture}
Fig. \ref{fig1} shows the proposed GAEI-UNet with a U-shaped encoder (left side)-decoder (right side) structure. Every step of the encoder includes a modified residual dropout block. The convolutional layer of each convolutional block is followed by a DropBlock, a batch normalization (BN) layer, and a rectified linear unit (ReLU), and then the convolution layer is utilized for downsampling with a stride size of 2. In each down-sampling step, we double the number of feature channels. Each step in the decoder includes a 2×2 transposed convolution operation for up-sampling and halves the number of feature channels, concatenates with the corresponding feature map from the encoder, which is then followed by the modified residual dropout block. The Attention-based Global context aggregator (AGCA) is added between the high-level part of the encoder and the decoder to enhance the global in-content information. At the final layer, a 1×1 convolution and Softmax activation function are used to get the output segmentation map.


\subsection{Attention-based Global context aggregator (AGCA)}
\begin{figure}[htbp]
    \centering
    \includegraphics[width=0.45 \textwidth]{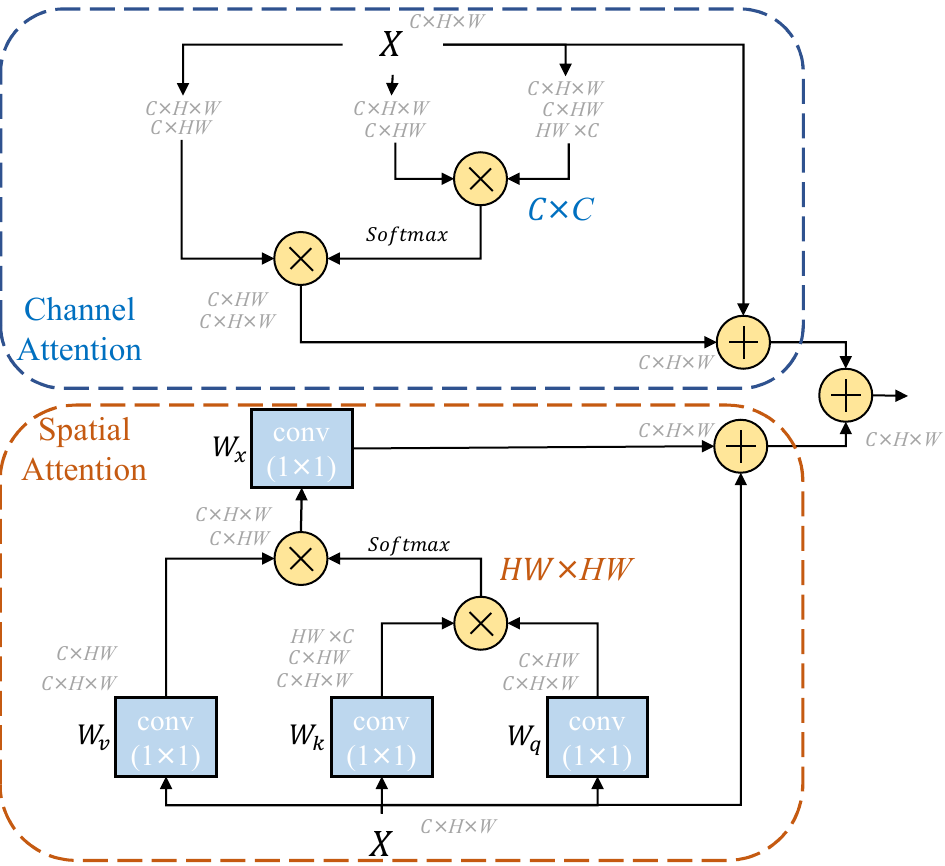}
    \caption{An overview of Attention-based Global Context Aggregator (AGCA)}
    \label{fig3}
\end{figure}

Attention-based Global context aggregator (AGCA) mainly consists of two parts -- two attention modules that capture long-range contextual information in spatial and channel dimensions respectively. It will explore more implicit features or relations for high-level semantics when it is inserted into the high-level U-net structure. The structure is shown in the Fig \ref{fig3}.
\subsubsection{Overview}
Given a picture for image segmentation, stuff or objects can be diverse in terms of scales, lighting, and views. Since convolution operations have a local receptive field, the features corresponding to pixels with the same label may have some differences. These differences can introduce intra-class inconsistency and affect recognition accuracy. To address this issue, we explore global contextual information by building associations among features using the attention mechanism. AGCA can adaptively aggregate long-range contextual information, thus improving feature representation for medical image segmentation. The whole block can be separated as two parallel parts -- spatial and channel attention global context aggregator and adopt element-wise add in the last.

\subsubsection{Spatial-attention global context aggregator}
As illustrated in the bottom part of Fig \ref{fig3}, the spatial-attention global context aggregator aims at strengthening the features of the query position via aggregating information from other positions.

We denote $\mathbf{x}={ \left\{ x_i\right\}}^{N_p}_{i=1}$ as the feature map of one input instance, such as an image or video, where $N_p$ is the number of positions in the feature map. For example, $N_p = H \times W$ for an image.
$\mathbf{x}$ and $\mathbf{y}$ represent the input and output of the aggregator, respectively with the same shape.

The spatial-attention global context aggregator can be expressed as
\begin{equation}
    \mathbf{y}_{i}=\mathbf{x}_{i}+W_{z} \sum_{j=1}^{N_{p}} \frac{f\left(\mathbf{x}_{i}, \mathbf{x}_{j}\right)}{\mathcal{C}(\mathbf{x})}\left(W_{v} \cdot \mathbf{x}_{j}\right)
\end{equation}
where $i$ is the index of query positions, and $j$ enumerates all possible positions. $f\left(\mathbf{x}_{i}, \mathbf{x}_{j}\right)$ denotes the relationship between position $i$ and $j$, and has a normalization factor $\mathcal{C}(\mathbf{x})$. $W_z$ and $W_v$ denote 1x1 convolution metrics. For simplification, we denote $\omega_{ij} = \frac{f\left(\mathbf{x}_{i}, \mathbf{x}_{j}\right)}{\mathcal{C}(\mathbf{x})}$ as normalized pairwise relationship between position $i$ and $j$.
We adopt $f$ in $\omega_{ij}$ as a dot-product similarity, formulated as 
\begin{equation}
    \omega_{i j}=\frac{\left\langle W_{q} \mathbf{x}_{i}, W_{k} \mathbf{x}_{j}\right\rangle}{N_{p}} = \frac{ \left \{ W_{q} \mathbf{x}_{i} \right \}^\top \cdot W_{k} \mathbf{x}_{j}}{N_{p}}
\end{equation}
with shape of $\mathbb{R}^{\left \{ H \times W \right \} \times \left \{ H \times W\right \}}$.

\subsubsection{Channel-attention global context aggregator}
Each channel map of high-level features can be regarded as a class-specific response, and different semantic responses are associated with each other. By exploiting the interdependencies between channel maps, we can emphasize interdependent feature maps and improve the feature representation of specific semantics. Therefore, we build a channel-attention global context aggregator to explicitly model interdependencies between channels, which is illustrated in the top part of Fig \ref{fig3}.

Silimar as spatial-attention global context aggregator, the channel-attention global context aggregator can be expressed as
\begin{equation}
    \mathbf{z}_{i}=\mathbf{x}_{i}+\sum_{j=1}^{C} \frac{f^\prime \left(\mathbf{x}_{i}, \mathbf{x}_{j}\right)}{\mathcal{C}(\mathbf{x})}\mathbf{x}_{j}
\end{equation}
where $i$ is the index of query positions, and $j$ enumerates all possible positions. $f\left(\mathbf{x}_{i}, \mathbf{x}_{j}\right)$ denotes the relationship between position $i$ and $j$, and has a normalization factor $\mathcal{C}(\mathbf{x})$. 1x1 convolution operations are ignored to keep the original relation between channels. For simplification, we denote $\omega_{ij}^\prime = \frac{f^\prime \left(\mathbf{x}_{i}, \mathbf{x}_{j}\right)}{\mathcal{C}(\mathbf{x})}$ as normalized pairwise relationship between position $i$ and $j$.
We adopt $f^\prime$ in $\omega_{ij}^\prime$ as a dot-product similarity, formulated as 
\begin{equation}
    \omega_{i j}^\prime=\frac{\left\langle \mathbf{x}_{i}, \mathbf{x}_{j}\right\rangle^\prime}{N_{p}} = \frac{ \mathbf{x}_{i} \cdot \left \{\mathbf{x}_{j} \right \}^\top }{N_{p}}
\end{equation}with shape of $\mathbb{R}^{C \times C}$.

\subsection{Elastic interaction-based loss function (EIL)}
The elastic interaction-based loss function is inspired by the elastic interaction between dislocations in crystalline defects \cite{anderson2017theory}, which is first introduced to medical image segmentation in \cite{elastic}. 

Consider two parameterized curves $\gamma_1$ and $\gamma_2$ in the $xy$ plane, which represents the boundary of two regions.
The total elastic energy is 
\begin{equation}
    \begin{aligned}
E& ={\frac{1}{8\pi}}\int_{\gamma_{1}\cup\gamma_{2}}\int_{\gamma_{1}^{\prime}\cup\gamma_{2}^{\prime}}{\frac{d\boldsymbol{l}\cdot d\boldsymbol{l}^{\prime}}{r}}  \\
&={\frac{1}{8\pi}}\int_{\gamma_{1}}\int_{\gamma_{1}^{\prime}}{\frac{d l_{1}\cdot d l_{1}^{\prime}}{r}}+{\frac{1}{8\pi}}\int_{\gamma_{2}}\int_{\gamma_{2}^{\prime}}{\frac{d l_{2}\cdot d l_{2}^{\prime}}{r}} \\
& +{\frac{1}{4\pi}}\int_{\gamma_{1}}\int_{\gamma_{2}}{\frac{d l_{1}\cdot d l_{2}}{r}}.
\end{aligned}
\end{equation}
, where $d l$ is the line element vector of the curve $\gamma$ and $r$ represents the distance between these corresponding two points.

There are two advantages of elastic energy. First, elastic interaction provides a strong attractive force to recombine a disconnected moving curve. This helps to maintain the continuity of the curve and prevent it from breaking apart. Second, the self-force of the curve has the effect of smoothing the moving curve. This helps to remove any small irregularities or noise in the curve and produce a smoother and more regular shape.

In \cite{elastic}, the simplified elastic interaction-based loss function is described as 
\begin{equation}
\begin{split}
    L_{elastic}=\frac{1}{8\pi}&\int_{\Omega}dxdy\int_{\Omega}\frac{\nabla\left(G_{t}+\alpha H\left(\phi\right)\right)\left(x,y\right)}{r} \cdot\\
&\frac{\nabla\left(G_{t}+\alpha H\left(\phi\right)\right)\left(x',y'\right)}{r}dx'dy' 
\end{split}
\label{loss}  
\end{equation}


, where $\phi$ is the level set representation of the moving curve $\gamma_2$, $\alpha$ is a hyperparameter and $H(\cdot)$ is a smoothing Heaviside function which controls the width of the contour by regularization parameter $\beta$. In order to reduce the computational complexity, we adopted the Fast Fourier Transform(FFT) to compute the loss function in Equation \ref{loss} and gradient efficiently in CNN \cite{elastic}.

\section{Experiment and Results}
\subsection{Datasets}

\textbf{DRIVE}: The Digital Retinal Images for Vessel Extraction (DRIVE) dataset \cite{staal2004ridge} is provided by a diabetic retinopathy screening program in the Netherlands. This dataset includes 40 images (3 × 565 × 584 pixels) which are divided into a training set and a test set, both containing 20 images and the corresponding ground truth. In DRIVE, two expert manual annotations are provided, the first of which is chosen as the ground truth for performance evaluation in the literature \cite{fu2016deepvessel}.
\subsection{Training Loss}
In order to stabilize the training process of EIL loss, we introduce a stabilizing term ($L_{CE}$) into the loss function.
Training loss consists of Binary-Cross-Entropy loss $L_{CE}$ and Elastic Interaction-Based Loss Function $L_{elastic}$. The overall loss function is defined as:
\begin{equation}
    Loss = \beta_1 L_{elastic} + \beta_2 L_{CE}
\end{equation}

In this project, we set $\beta_1=0.1$ and $\beta_2=1.5$. This setup allows the total loss to account for both local and global features, enabling the model to normalize the blood vessel features using global feature differences and prevent overfitting caused by excessive local features. By incorporating a long-range elastic interaction-based loss function, the model can capture global, long-range information for the thin, elongated structures in medical images. This approach enables the model to effectively balance local and global features, ensuring that the model captures the relevant features necessary for accurate segmentation while avoiding the negative effects of overfitting. 

\subsection{Implementation Details}
The GAEI-UNet model is trained from images using the augmented training set. The Adam optimizer is employed for both datasets. To keep the number of parameters small, the number of channels after the first convolutional layer is set to 16. The model is trained for 150 epochs, with a learning rate of 0.001 for the first 100 epochs and 0.0001 for the last 50 epochs. The size of the discard blocks of DropBlock is set to 7.
The batch size of the training is set to 8 and the dropout rate of DropBlock is set to 0.18. The implementation is based on Pytorch, and all experiments are run on an NVIDIA V100 GPU with 32 Gigabyte memory.
\subsection{Measurement metric}
To evaluate our model, we compare the segmentation results with the corresponding ground truth and divide the results of each pixel comparison into true positive (TP), false positive (FP), false negative (FN), and true negative (TN). We then compute sensitivity (SE), specificity (SP), F1 score (F1), and accuracy (ACC) evaluation metrics to compare the performance of the vessel detection, which are also calculated in Equation \ref{SE} \ref{SP} \ref{F1}. The area under the ROC curve (AUC) can also be used to measure the performance of the segmentation. A value of 1 for the AUC indicates perfect segmentation.

\begin{equation}
    SE = \frac{TP}{TP + FN}
    \label{SE}
\end{equation}
\begin{equation}
    SP = \frac{TN}{TN + FP}
    \label{SP}
\end{equation}
\begin{equation}
    F1 = \frac{2 \times (precision \times recall)}{precision + recall}
    \label{F1}
\end{equation}
where $precision = \frac{TP}{TP + FP}$ and $recall = \frac{TP}{TP + FN}$.

\subsection{Results}
\subsubsection{Comparisons With the U-Net Based Methods}
We conducted vessel segmentation experiments on the most popular networks, including U-Net \cite{unet}, UNet++\cite{zhou2019unet++}, AG U-Net\cite{zhang2019attention}, CS-Net\cite{mou2019cs}.

Table \ref{result} provides the qualitative results of vessel segmentation for retinal vessel dataset DRIVE. It is obviously apparent from these tables that GAEI-UNet is competitive with popular models by achieving the best overall performance with the highest AUC (0.9836), the highest SE(0.8346), the highest F1(0.8109), and the second SP(0.9838, only 0.1\% lower than the first AG U-Net\cite{zhang2019attention}) on DRIVE. Additionally, it is worth mentioning that the SE score obtained by GAEI-UNet is much higher than other methods on DRIVE since many researchers\cite{wu2021scs} have found that the more sensitivity, the more capable to segment thin vessels and boundary pixels, which demonstrates that GAEI-UNet has a great ability to extract microvascular structures.

\begin{table}[!ht]
    \centering
    \caption{Ablation Study on DRIVE dataset}
    \begin{tabular}{cccccccc}
    \hline
        Methods & SE & SP & AUC & F1 & ~   \\ \hline
        U-Net\cite{unet} & 0.7951 & 0.9812 & 0.9810 & 0.8053 & ~   \\ 
        U-Net++\cite{zhou2019unet++} & 0.7890 & 0.9813 & 0.9815 & 0.8067 & ~  \\ 
        AG U-Net\cite{zhang2019attention} & 0.8100 & \textbf{0.9848} & 0.9813 & - & ~   \\ 
        CS-Net\cite{mou2019cs} & 0.8112 & 0.9841 & 0.9801 & 0.8035 & ~   \\ 
        GAEI-UNet & \textbf{0.8346} & 0.9838 & \textbf{0.9836} & \textbf{0.8109} & ~  \\  \hline
    \end{tabular}
    \label{result}
\end{table}

\begin{figure*}[htbp]
    \centering
    \includegraphics[width=0.9 \textwidth]{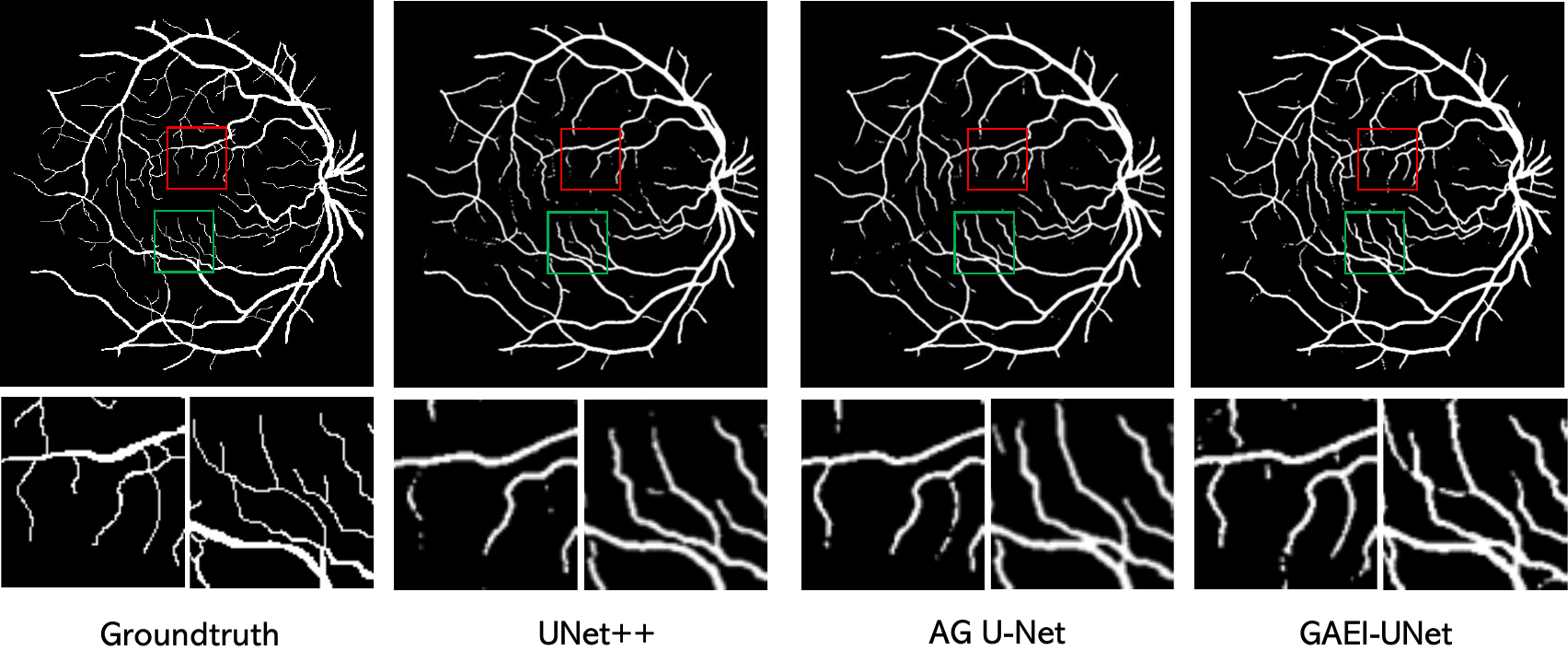}
    \caption{Segmentation results on \textbf{DRIVE}. From a horizontal perspective, the first column to the last column are the ground truths, and segmentation results of UNet++, AG U-Net, and GAEI-UNet, respectively. The DRIVE contains many thin and small vessel structures, and we zoom in on the image details for clearer visualization, as shown in the red and green box regions.} 
    \label{res}
\end{figure*}

We further visualize the vessel segmentation results, including those of UNet++, Attention U-Net, and FR-UNet, as shown in Fig. \ref{res}. The retinal vessel image contains many thin
vessels. Hence, we zoom in on the image details for clearer visualization, as shown in the highlighted rectangular regions. We can observe that GAEI-UNet detects more thin vessel pixels and more connected small structures than UNet++, AG U-Net.

\subsubsection{Ablation Study}
To evaluate the effectiveness of each component of the proposed GAEI-UNet model, ablation experiments were conducted on the DRIVE dataset.

\begin{table}[!ht]
    \centering
    \caption{Ablation Study on DRIVE dataset}
    \begin{tabular}{cccccccc}
    \hline
        Methods & SE & SP & AUC & F1 & ~   \\ \hline
        U-Net & 0.7951 & 0.9812 & 0.9810 & 0.8053 & ~   \\ 
        U-Net+Dropblock & 0.8137 & 0.9803 & 0.9837 & 0.8047 & ~  \\ 
        U-Net+AGCA & 0.8328 & 0.9821 & 0.9813 & 0.8090 & ~   \\ 
        GAEI-UNet & 0.8346 & 0.9838 & 0.9836 & 0.8109 & ~  \\  \hline
    \end{tabular}
    \label{ablation}
\end{table}

Table \ref{ablation} and Fig show the segmentation performance of U-Net, U-Net + Dropblock, U-Net + AGCA, and GAEI-UNet from top to bottom respectively.
From the results, we could obtain several useful observations: 
\begin{itemize}
    \item The GAEI-UNet model achieved better segmentation performance compared to the U-Net model. This indicates that the introduction of spatial and channel attention is an effective strategy for improving the accuracy of retinal vessel segmentation.
    \item The U-Net + AGCA model, which leverages an improved attention mechanism to enhance the extraction of high-level semantic features, demonstrated superior segmentation accuracy. This indicates that the AGCA module is effective in capturing more information from the deep feature map, leading to improved segmentation performance.
    \item The U-Net + Dropblock model, which addresses the issues of gradient vanishing and overfitting through the use of structured Dropout convolutional blocks, also achieved better segmentation performance compared to the original U-Net model. This suggests that the Dropblock technique is effective in preventing overfitting and improving the accuracy of blood vessel segmentation.
\end{itemize}
\section{Conclusion}
U-Net with the modified residual Dropout convolutional block we constructed has proven to be successful in addressing the issue of gradient vanishing and overfitting, eventually, leading to improved accuracy in blood vessel segmentation. 

Additionally, the proposed Attention-based Global context aggregator (AGCA) shows its ability to enhance the extraction of high-level semantic features, improving the overall segmentation performance.

Moreover, the combination of the long-range elastic interaction-based loss function (EIL) with the Cross-entropy Loss has been shown to effectively capture global, long-range information for thin structures such as blood vessels. This approach has proved to be particularly useful in medical imaging where the accurate segmentation of thin structures is essential for successful diagnosis and treatment.

Overall, our contributions have demonstrated significant improvements in blood vessel segmentation, and the proposed methodology can be applied to other medical image segmentation tasks. Our work has the potential to contribute to the development of more accurate and efficient medical image analysis tools, which can ultimately lead to better patient outcomes.

\vspace{12pt}
\bibliographystyle{IEEEtran}
\bibliography{IEEEabrv,output}

\begin{thebibliography}{10}
\providecommand{\url}[1]{#1}
\csname url@samestyle\endcsname
\providecommand{\newblock}{\relax}
\providecommand{\bibinfo}[2]{#2}
\providecommand{\BIBentrySTDinterwordspacing}{\spaceskip=0pt\relax}
\providecommand{\BIBentryALTinterwordstretchfactor}{4}
\providecommand{\BIBentryALTinterwordspacing}{\spaceskip=\fontdimen2\font plus
\BIBentryALTinterwordstretchfactor\fontdimen3\font minus
  \fontdimen4\font\relax}
\providecommand{\BIBforeignlanguage}[2]{{%
\expandafter\ifx\csname l@#1\endcsname\relax
\typeout{** WARNING: IEEEtran.bst: No hyphenation pattern has been}%
\typeout{** loaded for the language `#1'. Using the pattern for}%
\typeout{** the default language instead.}%
\else
\language=\csname l@#1\endcsname
\fi
#2}}
\providecommand{\BIBdecl}{\relax}
\BIBdecl

\bibitem{guo2014microfluidic}
Q.~Guo, S.~P. Duffy, K.~Matthews, A.~T. Santoso, M.~D. Scott, and H.~Ma,
  ``Microfluidic analysis of red blood cell deformability,'' \emph{Journal of
  biomechanics}, vol.~47, no.~8, pp. 1767--1776, 2014.

\bibitem{shen2016learning}
W.~Shen, M.~Zhou, F.~Yang, D.~Dong, C.~Yang, Y.~Zang, and J.~Tian, ``Learning
  from experts: developing transferable deep features for patient-level lung
  cancer prediction,'' in \emph{Medical Image Computing and Computer-Assisted
  Intervention--MICCAI 2016: 19th International Conference, Athens, Greece,
  October 17-21, 2016, Proceedings, Part II 19}.\hskip 1em plus 0.5em minus
  0.4em\relax Springer, 2016, pp. 124--131.

\bibitem{song2015lung}
J.~Song, C.~Yang, L.~Fan, K.~Wang, F.~Yang, S.~Liu, and J.~Tian, ``Lung lesion
  extraction using a toboggan based growing automatic segmentation approach,''
  \emph{IEEE transactions on medical imaging}, vol.~35, no.~1, pp. 337--353,
  2015.

\bibitem{lee2001automated}
Y.~Lee, T.~Hara, H.~Fujita, S.~Itoh, and T.~Ishigaki, ``Automated detection of
  pulmonary nodules in helical ct images based on an improved template-matching
  technique,'' \emph{IEEE Transactions on medical imaging}, vol.~20, no.~7, pp.
  595--604, 2001.

\bibitem{heimann2007statistical}
T.~Heimann, H.-P. Meinzer, and I.~Wolf, ``A statistical deformable model for
  the segmentation of liver ct volumes,'' \emph{3D Segmentation in the clinic:
  A grand challenge}, pp. 161--166, 2007.

\bibitem{acelastic}
Y.~Xiang, A.~C. Chung, and J.~Ye, ``An active contour model for image
  segmentation based on elastic interaction,'' \emph{Journal of Computational
  Physics}, vol. 219, no.~1, pp. 455--476, 2006.

\bibitem{liu2021review}
X.~Liu, L.~Song, S.~Liu, and Y.~Zhang, ``A review of deep-learning-based
  medical image segmentation methods,'' \emph{Sustainability}, vol.~13, no.~3,
  p. 1224, 2021.

\bibitem{ricci2007retinal}
E.~Ricci and R.~Perfetti, ``Retinal blood vessel segmentation using line
  operators and support vector classification,'' \emph{IEEE transactions on
  medical imaging}, vol.~26, no.~10, pp. 1357--1365, 2007.

\bibitem{moccia2018blood}
S.~Moccia, E.~De~Momi, S.~El~Hadji, and L.~S. Mattos, ``Blood vessel
  segmentation algorithms—review of methods, datasets and evaluation
  metrics,'' \emph{Computer methods and programs in biomedicine}, vol. 158, pp.
  71--91, 2018.

\bibitem{zhu2008detection}
X.~Zhu and R.~M. Rangayyan, ``Detection of the optic disc in images of the
  retina using the hough transform,'' in \emph{2008 30th annual international
  conference of the IEEE engineering in medicine and biology society}.\hskip
  1em plus 0.5em minus 0.4em\relax IEEE, 2008, pp. 3546--3549.

\bibitem{aquino2010detecting}
A.~Aquino, M.~E. Geg{\'u}ndez-Arias, and D.~Mar{\'\i}n, ``Detecting the optic
  disc boundary in digital fundus images using morphological, edge detection,
  and feature extraction techniques,'' \emph{IEEE transactions on medical
  imaging}, vol.~29, no.~11, pp. 1860--1869, 2010.

\bibitem{mihaylova2018spleen}
A.~Mihaylova and V.~Georgieva, ``Spleen segmentation in mri sequence images
  using template matching and active contours,'' \emph{Procedia Computer
  Science}, vol. 131, pp. 15--22, 2018.

\bibitem{chen2009automated}
W.~Chen, R.~Smith, S.-Y. Ji, K.~R. Ward, and K.~Najarian, ``Automated
  ventricular systems segmentation in brain ct images by combining low-level
  segmentation and high-level template matching,'' \emph{BMC medical
  informatics and decision making}, vol.~9, no.~1, pp. 1--14, 2009.

\bibitem{tsai2003shape}
A.~Tsai, A.~Yezzi, W.~Wells, C.~Tempany, D.~Tucker, A.~Fan, W.~E. Grimson, and
  A.~Willsky, ``A shape-based approach to the segmentation of medical imagery
  using level sets,'' \emph{IEEE transactions on medical imaging}, vol.~22,
  no.~2, pp. 137--154, 2003.

\bibitem{ATLI2021271}
\BIBentryALTinterwordspacing
İbrahim Atli and O.~S. Gedik, ``Sine-net: A fully convolutional deep learning
  architecture for retinal blood vessel segmentation,'' \emph{Engineering
  Science and Technology, an International Journal}, vol.~24, no.~2, pp.
  271--283, 2021. [Online]. Available:
  \url{https://www.sciencedirect.com/science/article/pii/S221509862030330X}
\BIBentrySTDinterwordspacing

\bibitem{unet}
O.~Ronneberger, P.~Fischer, and T.~Brox, ``U-{{Net}}: {{Convolutional
  Networks}} for {{Biomedical Image Segmentation}},'' 2015.

\bibitem{wang2019dual}
B.~Wang, S.~Qiu, and H.~He, ``Dual encoding u-net for retinal vessel
  segmentation,'' in \emph{Medical Image Computing and Computer Assisted
  Intervention--MICCAI 2019: 22nd International Conference, Shenzhen, China,
  October 13--17, 2019, Proceedings, Part I 22}.\hskip 1em plus 0.5em minus
  0.4em\relax Springer, 2019, pp. 84--92.

\bibitem{wu2019vessel}
Y.~Wu, Y.~Xia, Y.~Song, D.~Zhang, D.~Liu, C.~Zhang, and W.~Cai, ``Vessel-net:
  retinal vessel segmentation under multi-path supervision,'' in \emph{Medical
  Image Computing and Computer Assisted Intervention--MICCAI 2019: 22nd
  International Conference, Shenzhen, China, October 13--17, 2019, Proceedings,
  Part I 22}.\hskip 1em plus 0.5em minus 0.4em\relax Springer, 2019, pp.
  264--272.

\bibitem{zhang2019attention}
S.~Zhang, H.~Fu, Y.~Yan, Y.~Zhang, Q.~Wu, M.~Yang, M.~Tan, and Y.~Xu,
  ``Attention guided network for retinal image segmentation,'' in \emph{Medical
  Image Computing and Computer Assisted Intervention--MICCAI 2019: 22nd
  International Conference, Shenzhen, China, October 13--17, 2019, Proceedings,
  Part I 22}.\hskip 1em plus 0.5em minus 0.4em\relax Springer, 2019, pp.
  797--805.

\bibitem{su2022yolo}
Y.~Su, Q.~Liu, W.~Xie, and P.~Hu, ``Yolo-logo: A transformer-based yolo
  segmentation model for breast mass detection and segmentation in digital
  mammograms,'' \emph{Computer Methods and Programs in Biomedicine}, vol. 221,
  p. 106903, 2022.

\bibitem{zhou2017focal}
X.-Y. Zhou, M.~Shen, C.~Riga, G.-Z. Yang, and S.-L. Lee, ``Focal fcn: Towards
  small object segmentation with limited training data,'' \emph{arXiv preprint
  arXiv:1711.01506}, 2017.

\bibitem{tran2016fully}
P.~V. Tran, ``A fully convolutional neural network for cardiac segmentation in
  short-axis mri,'' \emph{arXiv preprint arXiv:1604.00494}, 2016.

\bibitem{anderson2017theory}
P.~M. Anderson, J.~P. Hirth, and J.~Lothe, \emph{Theory of dislocations}.\hskip
  1em plus 0.5em minus 0.4em\relax Cambridge University Press, 2017.

\bibitem{elastic}
Y.~Lan, Y.~Xiang, and L.~Zhang, ``An {{Elastic Interaction-Based Loss
  Function}} for {{Medical Image Segmentation}},'' in \emph{Medical {{Image
  Computing}} and {{Computer Assisted Intervention}} \textendash{} {{MICCAI}}
  2020}, A.~L. Martel, P.~Abolmaesumi, D.~Stoyanov, D.~Mateus, M.~A. Zuluaga,
  S.~K. Zhou, D.~Racoceanu, and L.~Joskowicz, Eds.\hskip 1em plus 0.5em minus
  0.4em\relax {Cham}: {Springer International Publishing}, 2020, vol. 12265,
  pp. 755--764.

\bibitem{staal2004ridge}
J.~Staal, M.~D. Abr{\`a}moff, M.~Niemeijer, M.~A. Viergever, and
  B.~Van~Ginneken, ``Ridge-based vessel segmentation in color images of the
  retina,'' \emph{IEEE transactions on medical imaging}, vol.~23, no.~4, pp.
  501--509, 2004.

\bibitem{fu2016deepvessel}
H.~Fu, Y.~Xu, S.~Lin, D.~W. Kee~Wong, and J.~Liu, ``Deepvessel: Retinal vessel
  segmentation via deep learning and conditional random field,'' in
  \emph{Medical Image Computing and Computer-Assisted Intervention--MICCAI
  2016: 19th International Conference, Athens, Greece, October 17-21, 2016,
  Proceedings, Part II 19}.\hskip 1em plus 0.5em minus 0.4em\relax Springer,
  2016, pp. 132--139.

\bibitem{zhou2019unet++}
Z.~Zhou, M.~M.~R. Siddiquee, N.~Tajbakhsh, and J.~Liang, ``Unet++: Redesigning
  skip connections to exploit multiscale features in image segmentation,''
  \emph{IEEE transactions on medical imaging}, vol.~39, no.~6, pp. 1856--1867,
  2019.

\bibitem{mou2019cs}
L.~Mou, Y.~Zhao, L.~Chen, J.~Cheng, Z.~Gu, H.~Hao, H.~Qi, Y.~Zheng, A.~Frangi,
  and J.~Liu, ``Cs-net: Channel and spatial attention network for curvilinear
  structure segmentation,'' in \emph{Medical Image Computing and Computer
  Assisted Intervention--MICCAI 2019: 22nd International Conference, Shenzhen,
  China, October 13--17, 2019, Proceedings, Part I 22}.\hskip 1em plus 0.5em
  minus 0.4em\relax Springer, 2019, pp. 721--730.

\bibitem{wu2021scs}
H.~Wu, W.~Wang, J.~Zhong, B.~Lei, Z.~Wen, and J.~Qin, ``Scs-net: A scale and
  context sensitive network for retinal vessel segmentation,'' \emph{Medical
  Image Analysis}, vol.~70, p. 102025, 2021.

\end{thebibliography}

\end{document}